\documentclass[a4paper]{jpconf}
\usepackage{graphicx}
\begin{document}
\title{Novel substrates for Helium adsorption: Graphane and Graphene--Fluoride}

\author{L Reatto$^1$, M Nava$^1$, D E Galli$^1$, C Billman$^2$, J O Sofo$^2$ and M W Cole$^2$}
\address{$^1$ Dipartimento di Fisica, Universit\`a degli Studi di Milano,
              Via Celoria 16, 20133 Milano, Italy}
\address{$^2$ Department of Physics and Materials Research Institute,
              Penn State University, University Park, PA 16802 USA}

\ead{luciano.reatto@mi.infn.it}

\begin{abstract}
The discovery of fullerenes has stimulated extensive exploration of the resulting 
behavior of adsorbed films.
Our study addresses the planar substrates graphene--fluoride (GF) and graphane (GH)
in comparison to graphene.
We present initial results concerning the potential energy, energy bands and low
density behavior of $^4$He and $^3$He films on such different surfaces.
For example, while graphene presents an adsorption potential that is qualitatively
similar to that on graphite, GF and GH yield potentials with different symmetry,
a number of adsorption sites double that on graphene/graphite and a larger 
corrugation for the adatom.
In the case of GF, the lowest energy band width is similar to that on graphite
but the He atom has a significantly larger effective mass
and the adsorption energy is about three time that on graphite.
Implications concerning the monolayer phase diagram of $^4$He are explored with the
exact path integral ground state method.
A commensurate ordered state similar to the $\sqrt{3}\times\sqrt{3}$ R30$^o$
state on graphite is found the be unstable both on GF and on GH.
The ground states of submonolayer $^4$He on both GF and GH are superfluids
with a Bose Einstein condensate fraction of about 10\%.
\end{abstract}

\section{Introduction}
Probably the best understood adsorption system is the He monolayer on graphite \cite{ref1}.
Experiments carried out at the University of Washington ca. 1970 revealed {\it for the 
first time} behavior corresponding to a two--dimensional (2D) gas. More dramatic was 
the appearance of a spectacular peak in the specific heat of $^4$He near $T_c=3$ K. 
This peak, well described by the 3 state Potts model, 
manifested a 2D transition from a high $T$ fluid to a low $T$ commensurate
($\sqrt{3}\times\sqrt{3}$ R30$^o$) phase, providing a benchmark measure of coverage, 
not seen in previous adsorption experiments. This ordered phase 
(at density $\Theta_{\sqrt{3}}=0.0636$\AA$^{-2}$) corresponds to atoms localized on second--nearest 
neighbor hexagons. At higher densities near completion of the first 
monolayer ($\Theta=0.11$\AA$^{-2}$) an incommensurate 2D triangular solid phase is present; 
the phase diagram at intermediate densities is not yet completely determined. 
A quantitative understanding of the He--graphite interaction was made possible by 
precise scattering measurements of surface bound states and band structures \cite{ref2}.

The availability of graphene (Gr) and its derivatives like graphane (GH) \cite{ref3} and 
graphene--fluoride (GF) \cite{ref4} offers the prospect of novel adsorption phenomena. 
No special phenomenon is expected for He adsorbed on one side of Gr because the 
interaction is similar to that on graphite. The situation is different for 
GH and GF due to the modified symmetry of the adsorption potential. 
We have developed a model adsorption potential for He on GF and GH. 
With exact Quantum Monte Carlo methods we have studied a single $^4$He and $^3$He 
atom on these substrates, as well as submonolayer films of  $^4$He at coverages 
similar to that ($\Theta=0.064$\AA$^{-2}$) of the $\sqrt{3}\times\sqrt{3}$ R30$^o$ state on graphite.

\section{Adsorption potential}
\begin{figure}[h]
\begin{minipage}{5cm}
\includegraphics*[width=9cm]{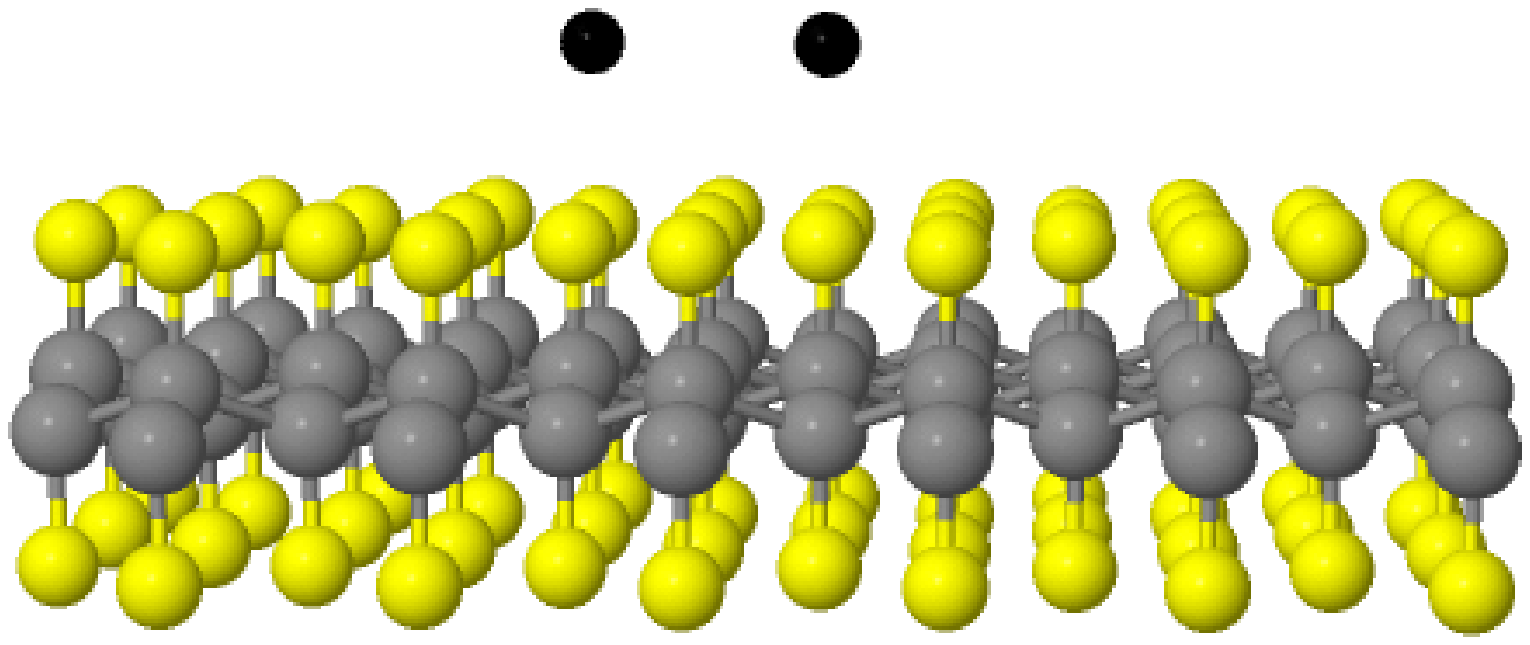}\label{a}
\end{minipage}\hspace{4cm}%
\begin{minipage}{5cm}
\includegraphics*[width=5cm]{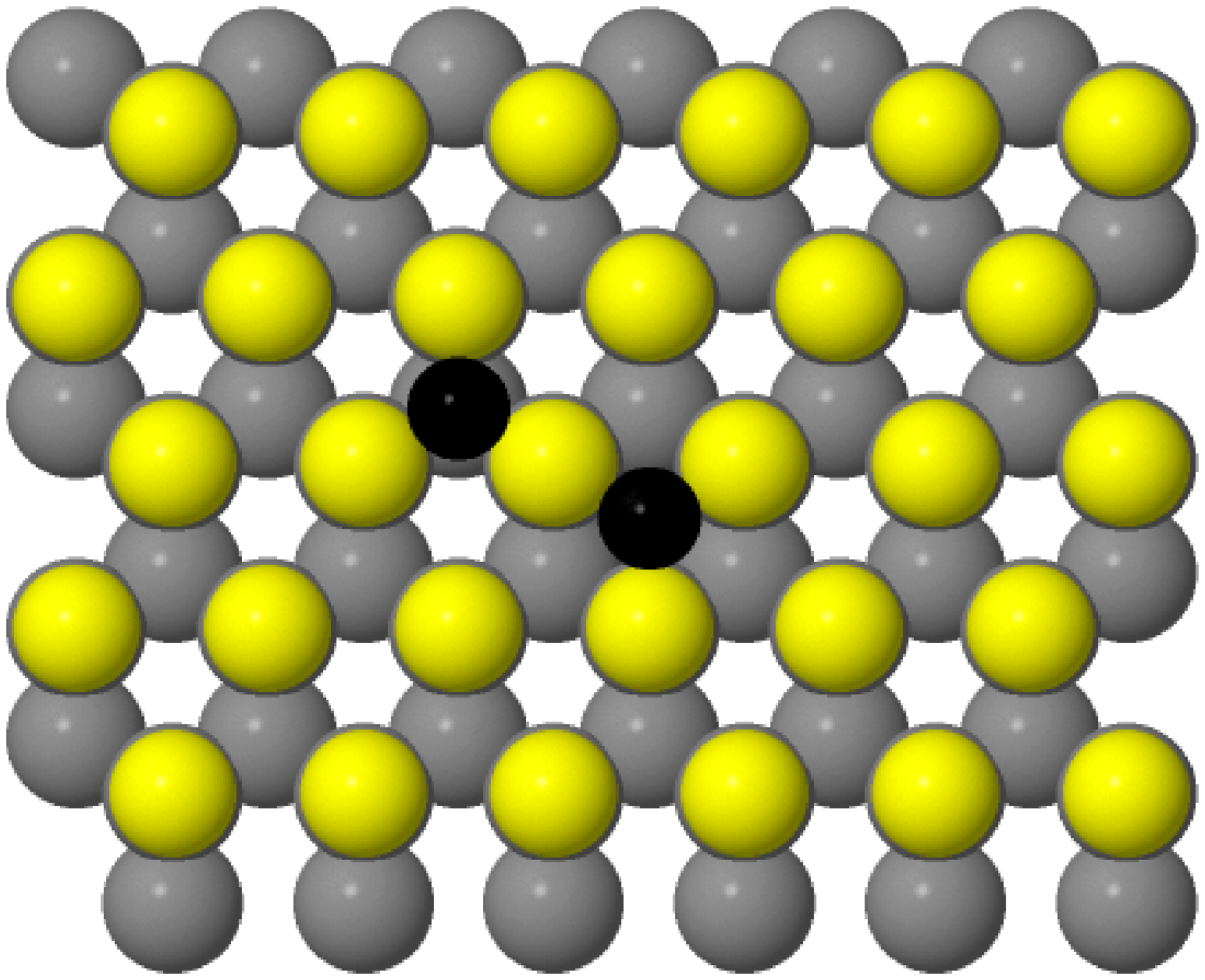}
\end{minipage}\caption{\label{fig1}
Two schematic views of GF. F (C) atoms are light (dark) gray. Positions of atoms
are to scale but their sizes are arbitrary. The black balls represent two
adsorption sites for He, one of each kind. GH is similar.
}
\end{figure}

Graphane and graphene--fluoride have a similar geometry; half of the H (F) 
atoms are attached on one side of the graphene sheet to the carbon atoms forming 
one of the two sublattices of graphene. 
The other half are attached on the other side to the C atoms forming the other sublattice. 
The H (F) atoms are located on two planes (see Fig.\ref{fig1}a); 
one is an overlayer located at a distance h above the pristine graphene plane 
while the other is an underlayer at a distance h below the graphene plane. 
In addition, there is a buckling of the C--plane with the C atoms of one 
sublattice moving upward by a distance b while the other sublattice moves 
downward by the same amount. A He atom approaching GH (GF) from above will 
interact primarily with the H (F) overlayer, but it will interact also with 
the C atoms and the H (F) atoms of the underlayer. 

We have adopted a traditional, semi--empirical model to construct the potential 
energy $V({\bf r})$ of a single He atom at position ${\bf r}$ near a surface \cite{ref5,ref6,ref7}. 
The potential is written $V({\bf r}) = V_{\rm rep}({\bf r})  + V_{\rm att}({\bf r})$, a sum of a Hartree--Fock 
repulsion derived from effective medium theory, and an attraction, $V_{\rm att}({\bf r})$, 
which is a sum of “damped” He atom van der Waals (VDW) interactions and the 
polarization interaction with the surface electric field. The first term 
is $V_{\rm rep}({\bf r})=\alpha \rho({\bf r})$.
Here $\alpha=364$ eV--bohr$^3$ is a value derived by several 
workers as the coefficient of proportionality between the repulsive 
interaction and the substrate's electronic charge density $\rho({\bf r})$ {\it prior} to adsorption. 
The geometry of GH and GF, their electronic charge density and the electrostatic 
potential have been obtained using Density Functional Theory with an all--electron 
triple numerical plus polarization basis set with an orbital cutoff of 3.7 \AA$\:$
as implemented in the DMol3 code \cite{ref8}. The exchange and correlation potential 
was treated in a Generalized Gradient Approximation parametrized by 
Perdew, Burke, and Ernzerhof \cite{ref9}. We use a tetragonal unit cell containing 
four C atoms and four H (F) atoms for GH and GF, respectively. 
The cell dimensions for GF are $a_1=2.59$
\AA, $a_2=4.48$
\AA, and $a_3=12$
\AA, while for GH we use $a_1=2.52$
\AA, $a_2=4.36$
\AA, and $a_3=12$
\AA. The Brillouin 
zone was sampled with a Monkhorst--Pack grid of $6\times3\times1$ {\bf k} points in both cases. 
The self--consistent cycles were run until the energy difference was less 
than 10$^{-6}$ eV. The atomic positions were relaxed until the forces on all 
atoms were lower than $0.01$
eV/\AA. As a result, the C--F distance is $1.38$
\AA, the C--C distance $1.57$
\AA, the C--C distance projected on the $x-y$ plane 
is $d=1.495$
\AA $\,$ and the buckling displacement $b=0.484$
\AA; while in GH, 
the C--H distance is $1.11$
\AA , the C--C distance $1.52$
\AA, $d=1.453$
\AA $\,$ and $b=0.45$
\AA. 

The attraction is a sum of contributions; for GH,
\begin{equation}
V_{\rm att}({\bf r}) = V_{\rm H+}({\bf r}) + V_{\rm gr}({\bf r}) + V_{\rm H-}({\bf r})
- \alpha_{\rm He} {\rm E}^{2}({\bf r})/2 
\end{equation}

The right--most term is the induced dipole energy, where $\alpha_{\rm He}=0.205$ \AA$^3$
is the static polarizability of the He atom and ${\bf E}({\bf r})$ is the electric field 
due to the substrate. The three VDW terms for GH originate from the H overlayer, 
the graphene sheet (we are neglecting in this term the small buckling of the 
graphene sheet) and the H underlayer, respectively. The graphene term may be 
written $V_{\rm gr}({\bf r}) =-A_C/z^4$, where $z$ is the He distance from the graphene sheet 
and $A_C = 3C_{3} d =1.84$
eV--\AA$^4$.
Here $C_3 =180$
meV--\AA$^3$
is the coefficient of 
proportionality entering the He--{\it graphite} VDW interaction and $d$ is the interlayer 
spacing of graphite. The term $V_{\rm H+}({\bf r})$, a sum of individual He--H interactions, 
requires the use of damping because the He--H separation can be small; 
we have adopted the Tang--Toennies damping procedure for this situation \cite{ref7} 
with the parameter
$\beta=3.78$ \AA$^{-1}$.
Also required are the VDW coefficients $C_{\rm 6H}$ 
entering the asymptotic He--H interaction, $-C_{\rm 6H}/r^6$.
Here we have used the value $C_{\rm 6H}=1.21$
eV--\AA$^6$, derived from the {\it ab initio} He--H2 interaction of Meyer, 
Hariharan and Kutzelnigg \cite{ref10}. $V_{\rm H-}({\bf r})$ is treated similarly to $V_{\rm H+}({\bf r})$
but it is considered to be function of $z$, with $A_{\rm H}=0.35$
eV--\AA$^4$,
because of the larger He distance from the H- plane ($\simeq 10$ \AA).      
$V_{\rm att}({\bf r})$ in the case of GF has an expression similar to (2) with $V_{\rm H+}$ and 
$V_{\rm H-}$ replaced by $V_{\rm F+}$ and by $V_{\rm F-}$ with the coefficient 
$C_{\rm 6F}=4.2$
eV--\AA$^6$
as given by Frigo et al \cite{ref11},
$\beta=3.2$
\AA$^{-1}$
and $A_{\rm F}=1.1$
eV--\AA$^4$.

With such model potentials the adsorption sites (see Fig.\ref{fig1}b) are above the 
centers of each triplets of H (F) atoms of the overlayer, forming a honeycomb 
lattice with the number of sites equal to the number of C atoms, twice as 
many as those on Gr. Half of the sites are above H (F) of the underlayer 
but the difference between the well depths for the two kinds of adsorption 
sites is very small, below 1\%.  For GF the well depth is 498 K and for 
GH it is 195 K. These values do not include the induced dipole energy which 
gives a contribution below 1\%. The inter--site energy barrier is 24 K for 
GF and 13 K for GH. Both values are significantly smaller than the barrier 
height 41K for graphite. In this last case the energy barrier does not 
depend much on the direction in the $x-y$ plane whereas in the case of GF and 
GH the ratio between maximum and minimum barrier height in the $x-y$ plane is 
of order of 4--5: the energy landscape of the two last substrates is 
characterized by a very large corrugation with narrow channels along which 
low potential barriers are present. The motion of the He atom, especially 
in the case of GF, visits only these channels , as though the atom moves 
in a multiconnected space. Another significant difference is that the 
distance between two neighboring sites is $1.49$
\AA $\,$ for GF and $1.45$
\AA $\,$ for GH whereas it is $2.46$
\AA $\,$ for graphite and for Gr.
Prior to these studies, graphite was believed to be the most attractive 
surface for He, with a well--depth a factor of 10 greater than that on the 
least attractive surface (Cs). If correct, the present results reveal GF 
to replace graphite, since its well is a factor of 3 more attractive. 

\section{A single Helium atom on the substrates}
We verified that the electrostatic contribution to $V_{\rm att}$ is small so we 
neglected it in this first computation.  We computed the exact ground 
state energy of one $^4$He atom or one $^3$He atom on GF and GH, see 
Table \ref{table1}. The binding energy on GH is similar to that on graphite, 
whereas that on GF is about three times that on graphite.  In both cases 
the ground state is delocalized over the full substrate and both kinds 
of adsorption sites are occupied with comparable probability. 

Our computation is based on the Path Integral Ground State (PIGS) method \cite{ref12}.  
With this method we can compute quantum averages of the ground state of 
the system using the quantum evolution in imaginary--time $\tau$ of a trial 
wave function $\Psi_t$. If $\Psi_t$ is not orthogonal to the ground state, and $\tau$
is sufficiently long, the quantum evolution purges from $\Psi_t$ the 
contributions of the excited states, yielding the ground state energy and 
wave function. A valuable feature of the PIGS method is that it is exact, 
in principle; the results are independent of $\Psi_t$ \cite{ref13} and systematic errors 
may be reduced below the statistical uncertainty. 
An additional feature 
of PIGS is that it can be used for a single particle as well as for many--body 
Bose systems like the system in the study reported in sect. 3. 

\begin{figure}[h]
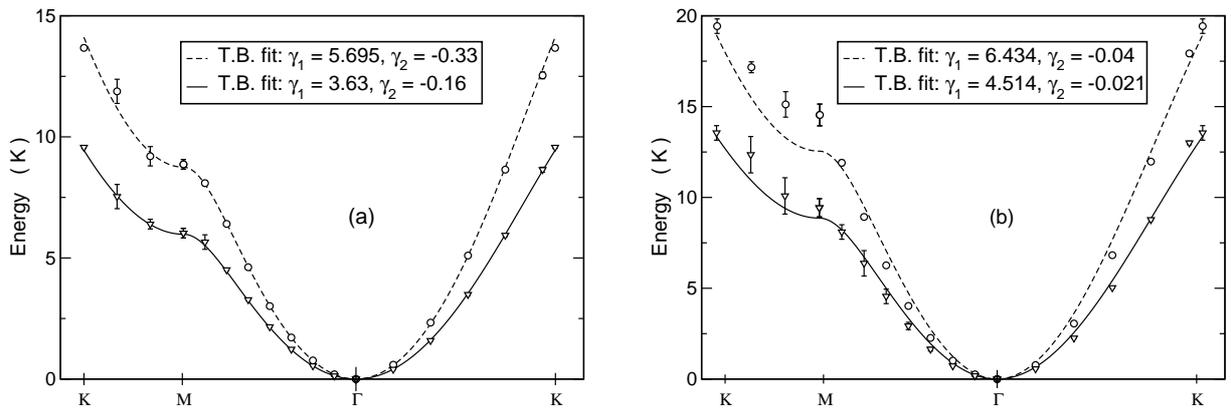

\begin{minipage}{18pc}
\includegraphics*[width=18pc]{fig2a.eps}\label{a}
\end{minipage}\hspace{2pc}%
\begin{minipage}{18pc}
\includegraphics*[width=18pc]{fig2b.eps}
\end{minipage}\caption{\label{fig2}
Panel (a): The lowest energy band of $^3$He and $^4$He on GF along the directions
$\Gamma$K, $\Gamma$M and MK.
Error bars represent uncertainty from the inversion procedure. 
Lines (dashed line for $^3$He and full line for $^4$He) are the fit with the tight
binding model with parameters in the inset.
Panel (b): Same as in panel (a) on GH. 
}
\end{figure}

With the PIGS method one can compute also imaginary--time correlation 
functions like the density--density one in Fourier space at imaginary 
time $\tau$:
$S({\bf k},\tau) = \langle \rho_{\bf k}(\tau) \rho_{-{\bf k}}(0) \rangle$,
$\rho_{\bf k}(\tau) = \exp \left[ i {\bf k} \cdot \hat{\bf r}(\tau)\right]$.
Here $\hat{\bf r}(\tau)$ is the position of the 
atom at imaginary time $\tau$.
$S({\bf k},\tau)$ contains  information  on the excited  
states of the system. The state of wave vector ${\bf k}$ of the lowest energy 
band has a dominant role in $S({\bf k},\tau)$ and its energy can be estimated via 
Laplace inversion using powerful inversion methods \cite{ref14}. The computed 
energy spectrum along the directions $\Gamma$K and $\Gamma$M for He on GF and on 
GH is shown in Fig.\ref{fig2}.
These bands are represented rather accurately by a tight binding model with
nearest and next nearest coupling \cite{reftb}.
For comparison we have computed with this same 
method the band energy for He on graphite finding substantial agreement 
with the Carlos and Cole result for the lowest band \cite{ref17}. 
The bandwidths $\Delta$ of He on these three substrates are given in Table \ref{table1}.

\begin{table}[h]
\caption{\label{table1}
Kinetic, potential and total energies for the ground state of He on GF, 
on GH and on graphite. In the last column the bandwidth $\Delta$ is shown. 
Numbers in parentheses represent statistical uncertainty in the last digit.
}

\begin{center}
\lineup
\begin{tabular}{*{7}{l}}
\br
System  &$E_{kin}$ (K)& $E_{pot}$ (K)&$E_{tot}$ (K)&$\Delta$ (K) \cr
\mr
$^4$He+GF & 46.78(2) & -422.94(1) & -376.16(3) &  9.6(1) \cr
$^3$He+GF & 51.08(1) & -413.41(1) & -362.33(1) & 13.7(1) \cr
\mr
$^4$He+GH & 20.51(1) & -153.58(1) & -133.07(2) & 13.6(4) \cr
$^3$He+GH & 22.53(1) & -149.50(1) & -126.97(3) & 19.4(4) \cr
\mr
$^4$He+Gr & 25.30(4) & -168.49(1) & -143.19(4) &  9.6(2) \cr
$^3$He+Gr & 27.05(2) & -162.87(1) & -135.82(2) & 15.7(4) \cr
\br
\end{tabular}
\end{center}
\end{table}

The effective masses $m^{\star}$ of the various systems reflect the varying 
corrugations of the potentials. For $^4$He ($^3$He), the ratios 
of $m^{\star}$ to the bare mass are 1.40 (1.25), 1.10 (1.08) and 1.05 (1.01) 
on GF, graphite and GH, respectively. The smaller mass enhancement 
of $^3$He than $^4$He reflects the smaller ratio of the corrugation 
potential to the translational zero--point energy.

\section{Ground state of submonolayer $^4$He on GF}
\begin{figure}[h]
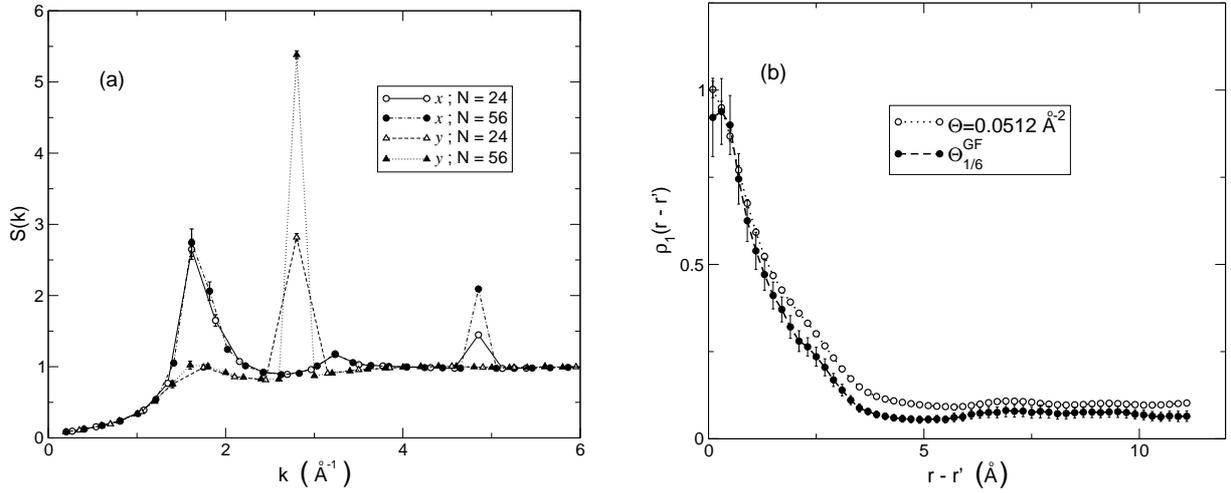

\begin{minipage}{18pc}
\includegraphics*[width=18pc]{fig3a.eps}\label{a}
\end{minipage}\hspace{2pc}%
\begin{minipage}{18pc}
\includegraphics*[width=18pc]{fig3b.eps}
\end{minipage}\caption{\label{fig3}
Static structure factor at density $\Theta^{\rm GF}_{1/6}$ (panel (a))
and off diagonal one body density
matrix (panel (b)) for $^4$He on GF at $\Theta=0.0512$ \AA$^{-2}$ (open circles) and
$\Theta^{\rm GF}_{1/6}=0.0574$ \AA$^{-2}$ (filled circles).
Lines are guides to the eyes.
}
\end{figure}
We have studied a $^4$He submonolayer on GF. As He--He interaction we 
have used an Aziz potential \cite{ref15}. The ground state has been computed 
for a number of $^4$He atoms from 22 to about 100 spanning the density 
range $\Theta$=0.04--0.09 \AA$^{-2}$.
On graphite the ground state is the commensurate
$\sqrt{3}\times\sqrt{3}$ R30$^o$ state with filling factor 1/3 of the 
adsorption sites. A similar state on GF is obtained by populating 
fourth neighbor sites (this corresponds to second neighbors in one 
of the sublattices of the honeycomb at a distance 4.482 \AA) 
with a filling factor of the adsorption sites equal to $1/6$
and it corresponds to a density $\Theta^{\rm GF}_{1/6}=0.0574$ \AA$^{-2}$.
Notice that this density is smaller than the 
$\Theta_{\sqrt{3}}=0.0636$ \AA$^{-2}$
on graphite due to the dilation of the C plane in GF. 
We find that this ordered 
state is unstable: starting the simulation from an ordered configuration 
after a short Monte Carlo evolution the Bragg peaks corresponding to 
the $\sqrt{3}\times\sqrt{3}$ R30$^o$ state disappear and the system 
evolves into a disordered fluid state modulated by the substrate 
potential. $S(k)$ at this density is plotted in Fig.\ref{fig3}a as function of 
$k_x$ and $k_y$ for two numbers $N$ of particles: the intensity of some of 
the peaks do not depend on $N$ so they are due to short range order, 
others scale roughly as $N$ and arise from the modulation of the density 
due to the adsorption potential.
The energy has a minimum value $E_0=-377.71 \pm 0.04$ K per atom at the 
density $\Theta_{eq}=0.049$ \AA$^{-2}$.
This lies 1.5 K below the single particle 
energy, implying that the ground state is a self--bound liquid. 
For comparison, we note that the strictly 2D cohesive energy of 
$^4$He \cite{ref16} is just 0.84 K and the equilibrium density is $\Theta=0.0436$ \AA$^{-2}$. 

We have computed the off diagonal one body density matrix $\rho_1(r-r')$.
As can be seen in Fig.\ref{fig3}b $\rho_1$ reaches a plateau at large $r-r'$ and the 
Bose Einstein condensate (BEC) fraction is $10.3 \pm 0.4$ \% at $\Theta=0.0512$ \AA$^{-2}$
and $7.3 \pm 1.5$ \% at $\Theta_{1/6}$; the system is superfluid.  
We reach a similar conclusion in the case of the GH substrate: 
the ground state is a liquid with density $\Theta_{eq}=0.042$ \AA$^{-2}$ 
and $E_0=-134.02 \pm 0.05$ K per atom and the BEC fraction is $22.6 \pm 1.3$ \% near the 
equilibrium density and $6.8 \pm 0.5$ \% at $\Theta^{\rm GH}_{1/6}=0.0608$ \AA$^{-2}$. Note that this condensate 
fraction is significantly smaller than the value ($\simeq 40$ \%) for $^4$He 
in 2D \cite{ref16}. The smaller value is a consequence of the spatial order, 
albeit imperfect, induced by the substrate potential and of the 
smaller effective surface available to the atoms due to the strong 
channeling induced by that potential.

\section{Discussion}
He adsorption on new substrate materials is valuable because of the 
fundamental importance of helium in many--body physics, with a variety of 
phases seen in both 2D and 3D. Our results indicate that the GF substrate 
provides the strongest binding of any surface (since the previous record 
was held by graphite). Moreover, the novel symmetry, the smaller 
intersite distance and large corrugation imply that quite novel properties 
may be anticipated for this system. This is indeed the case. 
When many $^4$He atoms are adsorbed on GF and on GH the most striking result 
is that the ground state is a low density liquid modulated by the substrate 
potential and the system has BEC, i.e. it is a superfluid. 
This is qualitatively different from graphite for which the lowest energy 
state is the $\sqrt{3}\times\sqrt{3}$ R30$^o$ commensurate one with no BEC \cite{ref18}. 
We have verified that such an ordered state on GF and GH is unstable 
relative to the liquid phase.  In subsequent work we will provide predictions 
concerning the phase diagrams and thermodynamic properties for both He/GF and He/GH,
hoping to stimulate experimental studies of these systems. 

It should be noticed that some of the parameters in the adsorption potential 
are not known with high precision or they have been adopted from other systems. 
We have verified that even a change of parameters like $\alpha$, $C_{\rm 6F}$ and $\beta$
by 10--20\% does not modify the qualitative behavior of the adsorbed He atoms 
even if there can be a sizable change in the value, for instance, of the 
adsorption energy.
Measurement of thermodynamic properties and 
He atomic beam scattering experiments from GF and GH will be 
important to test the accuracy of our model potentials.

\section*{Acknowledgements}
This work has been supported by Regione Lombardia and CILEA Consortium through a LISA
Initiative (Laboratory for Interdisciplinary Advanced Simulation) 2010 grant 
[http://lisa.cilea.it]. 
Chris Billman and Jorge Sofo are partially supported by the Donors of
the Petroleum Research Fund administrated by the American Chemical
Society.

\end{document}